\documentclass[runningheads]{llncs}
\usepackage{graphicx}
\usepackage{amsfonts}
\usepackage{amsmath}
\usepackage{subfig}

\usepackage{float}
\usepackage[ruled]{algorithm2e}

\usepackage[subtle,margins=normal,leading=normal]{savetrees} 

\newcommand{\HH}{\ensuremath{\mathbf{H}}}

\newcommand{\V}{\ensuremath{\mathbf{V}}}
\newcommand{\W}{\ensuremath{\mathbf{W}}}

\renewcommand{\aa}{\ensuremath{\mathbf{a}}}

\renewcommand{\c}{\ensuremath{\mathbf{c}}}

\newcommand{\h}{\ensuremath{\mathbf{h}}}

\newcommand{\w}{\ensuremath{\mathbf{w}}}

\newcommand{\z}{\ensuremath{\mathbf{z}}}

\begin{document}


\title{Accounting~for~Dependencies~in~Deep Learning~based~Multiple~Instance~Learning for Whole Slide Imaging}

\author{Andriy Myronenko\orcidID{0000-0001-8713-7031} \and
Ziyue Xu\orcidID{0000-0002-5728-6869} \and
Dong Yang\orcidID{0000-0002-5031-4337} \and
Holger Roth\orcidID{0000-0002-3662-8743} \and
Daguang Xu\orcidID{0000-0002-4621-881X} 
}

\authorrunning{A. Myronenko et al.}
%
\institute{NVIDIA, Santa Clara, CA, USA \\
\email{amyronenko@nvidia.com}}

\maketitle              
\begin{abstract}
Multiple instance learning (MIL) is a key algorithm for classification of whole slide images (WSI). Histology WSIs can have billions of pixels, which create enormous computational and annotation challenges. Typically, such images are divided into a set of patches (a bag of instances), where only bag-level class labels are provided. Deep learning based MIL methods calculate instance features using convolutional neural network (CNN). Our proposed approach is also deep learning based, with the following two contributions: Firstly, we propose to explicitly account for dependencies between instances during training by embedding self-attention Transformer blocks to capture dependencies between instances. For example, a tumor grade may depend on the presence of several particular patterns at different locations in WSI, which requires to account for dependencies between patches. Secondly, we propose an instance-wise loss function based on instance pseudo-labels. We compare the proposed algorithm to multiple baseline methods, evaluate it on the PANDA challenge dataset, the largest publicly available WSI dataset with over 11K images, and demonstrate state-of-the-art results. 

\keywords{multiple instance learning, histopathology, transformer, whole slide imaging, self-attention}
\end{abstract}

\section{Introduction}
\begin{figure}[t]
  \centering
    \includegraphics[width=0.7\textwidth]{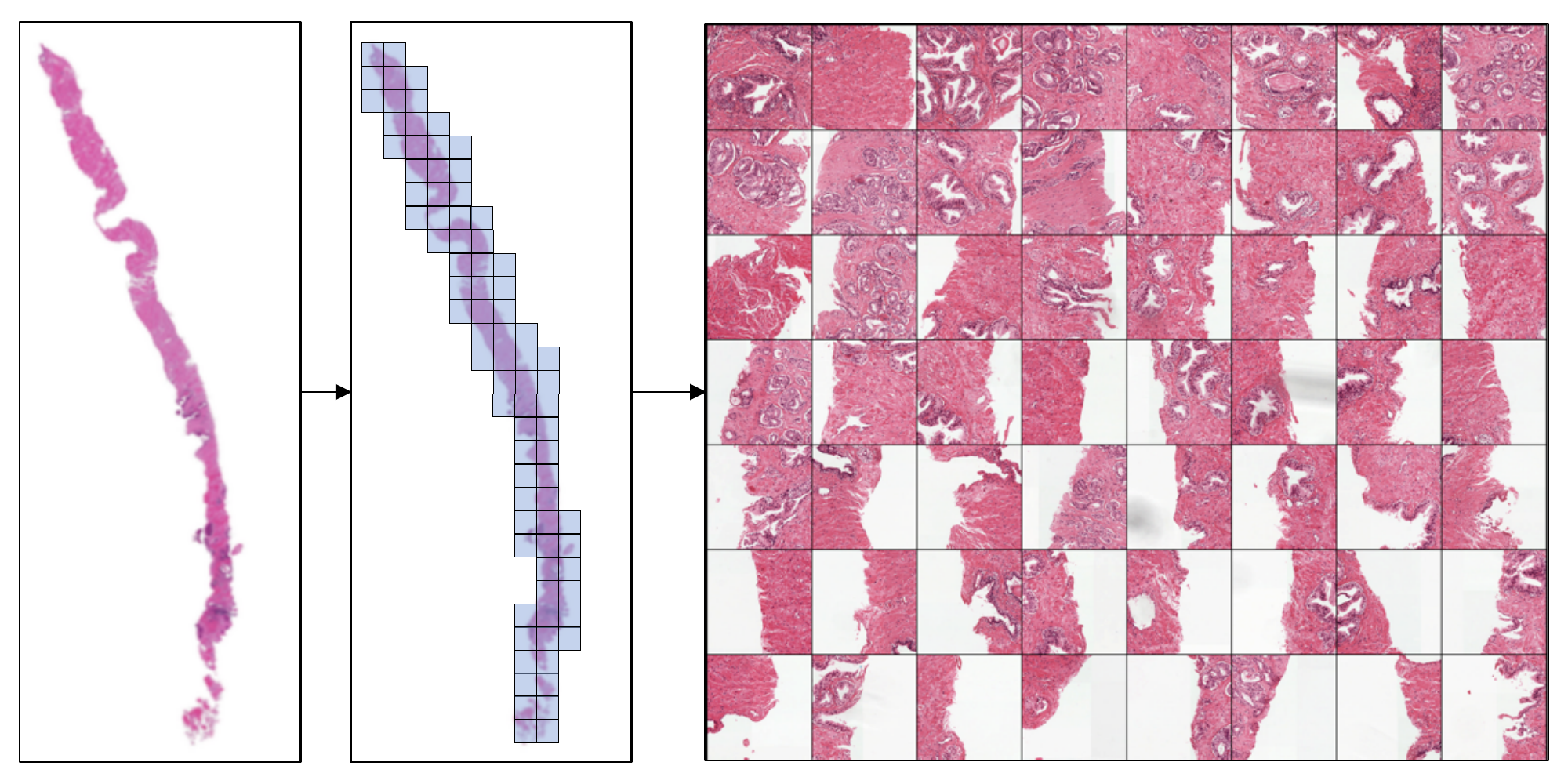}
  \caption{An example of patch extraction from WSI from the PANDA challenge dataset~\cite{Panda2020}. We tile the image and retain only the foreground patches, out of which we take a random subset to form a bag.}
  \label{fig:patches}
\end{figure}

Whole slide images (WSI) are digitizing histology slides often analysed for diagnosis of cancer~\cite{Campanella19}. WSI can contain several billions pixels, and are commonly tiled into smaller patches for processing to reduce the computational burden (Figure~\ref{fig:patches}). Another reason to use patches is because the area of interest (tumor cells) occupies only a tiny fraction of the image, which impedes the performance of conventional classifiers, most of which assume that the class object occupies a large central part of the image. Unfortunately, patch-wise labels are usually not available, since the detailed annotations are too costly and time-consuming. An alternative to supervised learning is weakly-supervised learning, where only a single label per WSI is available.

Multiple Instance Learning (MIL) is a weakly supervised learning algorithms, which aims to train a model using a set of weakly labeled data~\cite{Dietterich97,Maron98}. Usually a single class label is provided for a bag of many unlabeled instances, indicating that at least one instance has the provided class label. It has many applications in computer vision and language processing~\cite{Carbonneau18}, however learning from bags raises important challenges that are unique to MIL. In  context of histopathology, a WSI represents a bag, and the extracted  patches (or their features) represent instances (we often use these notations interchangeably). 

With the advent of convolutional neural networks (CNN), deep learning based MIL has become the mainstream methodological choice for WSI~\cite{Srinidhi21}. Campanella et al.~\cite{Campanella19} was one of the first works to conduct a large study on over 44K WSI,
laying the foundation for MIL applications in clinical practise. Since the instance labels are not known, classical MIL algorithm usually selects only one (or a few) instances based on the maximum of the prediction probability at the current iteration.  Such approach is very time consuming, as all patches need to be inferenced, but only a single patch contributes to the training of CNNs at each iteration. Ilse et al.~\cite{Ilse18} proposed to use an attention mechanism (a learnable weights per instance) to utilize all image patches, which we also adopt. 

More recent MIL methods include works by Zhao et al.~\cite{Zhao20}, who proposed to pre-train a feature extractor based on the variational auto-encoder, and use a graph convolutional  network for final classification. 

Hashimoto et al.~\cite{Hashimoto20} proposed to combine MIL with domain adverserial normalization and multiple scale learning. 
Lu et al.~\cite{Lu21} precomputed patch-level features (using pretrained CNN) offline to speed up training, and proposed an additional clustering-based loss to improve generalization during MIL training.
Maksoud et al.~\cite{Maksoud20} proposed  to use a hierarchical approach to process the  down-scaled WSI first, followed by by high resolution processing when necessary. Such approach demonstrated significant reduction in processing time, while maintaining the baseline accuracy.

We observed that most MIL methods assume no dependencies among instances, which is seldom true especially in histopathology~\cite{Srinidhi21}. Furthermore, a lack of instance-level loss supervision creates more opportunities for CNNs to overfit.

In this work, we propose a deep learning based MIL algorithm for WSI classification with the following contributions:
\begin{itemize}
    \item we propose to explicitly account for dependencies between instances during training. We embed transformer encoder~\cite{Vaswani17} blocks into the classification CNN to capture the dependencies between instances. 

    \item we propose an instance-wise loss supervision based on instance pseudo-labels. The pseudo-labels are computed based on the ensemble of several models, by aggregating the attention weights and instance-level  predictions.
\end{itemize}

 We evaluate the proposed method on PANDA challenge~\cite{Panda2020} dataset, which is currently the largest publicly available WSI dataset with over 11000 images, against the baseline methods as well as against the Kaggle challenge leaderboard with over 1000 competing teams, and demonstrate state-of-the-art (SOTA) classification results.

\section{Method}
\label{sec:method}

MIL aims to classify a bag of instances $\HH = \{\h_1,\dots,\h_K\}$ as positive if at least one of the instances $\h_k$ is positive. The number of instances $K$  could vary between the bags. Individual instance labels are unknown, and only the bag level label $Y=[0,1]$ is provided:

\begin{equation}
Y = 
\begin{cases} 
0 ,& \text{iff  all  }  y_{k} = 0, \\
1 ,&\text{iff  any }  y_{k} = 1.
\end{cases}
\end{equation}
which is equivalent to $Y = \max_{k} \{ y_{k} \}$ definition using a Max operator. 

Training a model whose loss is based on the maximum over instance labels is problematic due to vanishing gradients~\cite{Ilse18}, and the training process becomes slow since only a single patch contributes to the optimization. Ilse et al.~\cite{Ilse18} proposed to use all image patches as linear combination weighted by attention weights. Consider 
$\HH \in \mathbb{R}^{M\times K}$ to be instance embeddings, e.g features of a CNN final layer after average pooling. Then a linear combination of patch embeddings is 

\begin{equation}
\label{eq:weighted_sum}
\z = \sum_{k=1}^{K} a_k \h_k =  \HH \aa
\end{equation}
where the attention weights of patch embeddings are $\aa=\textit{softmax}\big{(} \tanh \big{(} \HH\V)\w\big{)}$,

where $\w \in \mathbb{R}^{L\times 1}$ and $\V \in \mathbb{R}^{M \times L}$ are parameters. The attention weights are computed using a multilayer perceptron (MLP) network with a single hidden layer.

\subsection{Dependency between instances}

The assumption of no dependency between the bag instances often does not hold. For example, for grading the severity of prostate cancer, pathologists need to find two distinct tumor growth patterns in the image and assign Gleason scores to each~\cite{Bulten21}. Then the International Society of Urological Pathology (ISUP) grade is calculated, based on the combination of major and minor Gleason patterns. ISUP grade indicates a severity of the tumor and plays a crucial role in treatment planning. Here, we propose to use the self-attention to account for dependencies between instances. 
In particular, we adopt the transformer, which was initially introduced to capture long range dependencies between words in sentences~\cite{Vaswani17} and later applied to vision~\cite{Dosovitskiy21}. Whereas traditional convolutions are local operation, the self-attention block of Transformers computes attention between all combinations of tokens at a larger range directly.

\begin{figure}[t]
  \centering
    \includegraphics[width=1.0\textwidth]{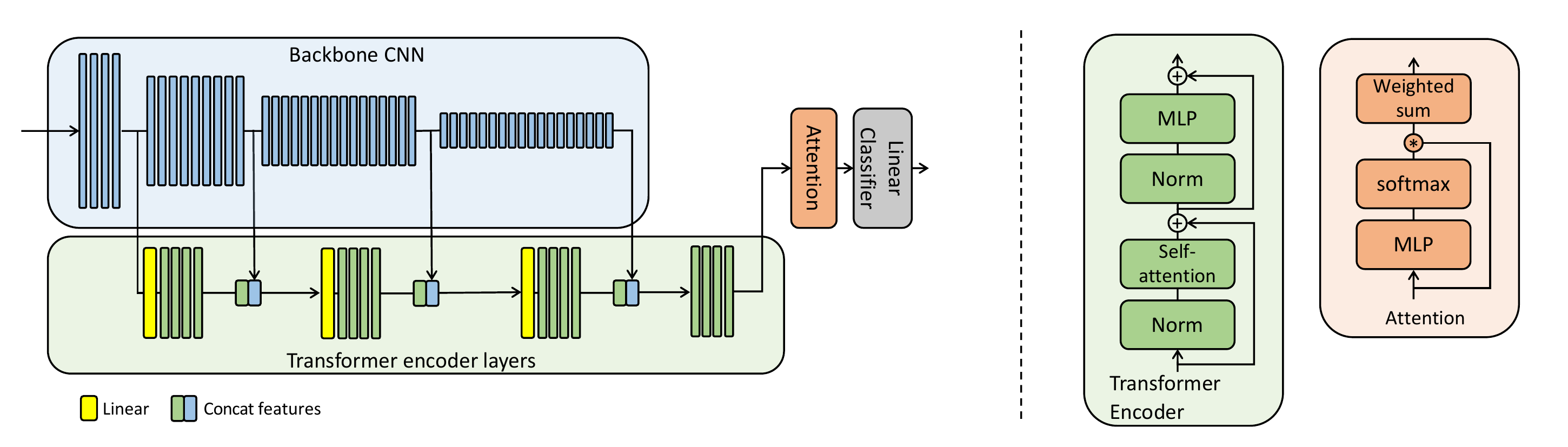}
  \caption{Model architecture overview. The backbone CNN (blue) extracts features at different scales, which are spatially averaged-pooled before feeding into the transformer encoder layers (green), to account for dependencies between instances.  The input to the network is $B\times N\times 3\times W\times H$. Where $B$ is the batch size, $N$ is the number of instances (patches extracted from a single whole slide image), and $3\times W\times H$ is the spatial patch size.}
  \label{fig:net}
\end{figure}

A key component of transformer blocks is a scaled dot product self-attention which is defined as $softmax(QK^T  / \sqrt{d})V$, where queries $Q$, keys $K$, and values $V$ matrices are all derived as linear transformations of the input (in our case the instance features space $\HH$).
The self-attention is performed several times  with different, learned linear projections in parallel (multi-head attention). 
In addition to self-attention, each of the transformer encoder layers also contains a fully connected feed-forward network and layer normalization (see Figure~\ref{fig:net})~\cite{Vaswani17,Dosovitskiy21}. 

We propose two variants of utilizing transformers. In the simplest case we attach a transformer encoder block only to the end of the backbone classification CNN after avg pooling. The idea is similar to the approach proposed in Visual transformers, but before avg pooling~\cite{Dosovitskiy21}.
The difference here is that in Visual transformers, the goal was to account for dependencies between the spatial regions (16px$\times$16px) of the same patch. Whereas we want to account for the dependencies among the patches.  Another relevant work was proposed by Wang et al.~\cite{Wang19} to utilize self-attention within MIL, but for text-based disease symptoms classification.
We maintain the dimensionality of encoded data, so that the input, output and hidden dimensionality of the transformer encoder are the same. We  call it Transformer MIL.

We also consider a variant of a deeper integration of the transformer with the backbone CNN. We attach separate transformer encoder blocks after each of the main ResNet blocks~\cite{He16} to capture the patch encodings at different levels of its feature pyramid. The output of the first transformer encoder is concatenated with next feature scale space of ResNet (after average pooling), and is fed into the next level transformer encoder, up until the final encoder layer, followed by the attention layer. We want to capture dependencies between patches at multiple scales, since different level of CNN output features include different semantic information. Such a Pyramid Transformer MIL network is shown in Figure~\ref{fig:net}.

\subsection{Instance level semi-supervision and pseudo-labeling}

\begin{figure}[ht]
    \centering
    \subfloat[]{\includegraphics[angle=90,width=0.5\textwidth]{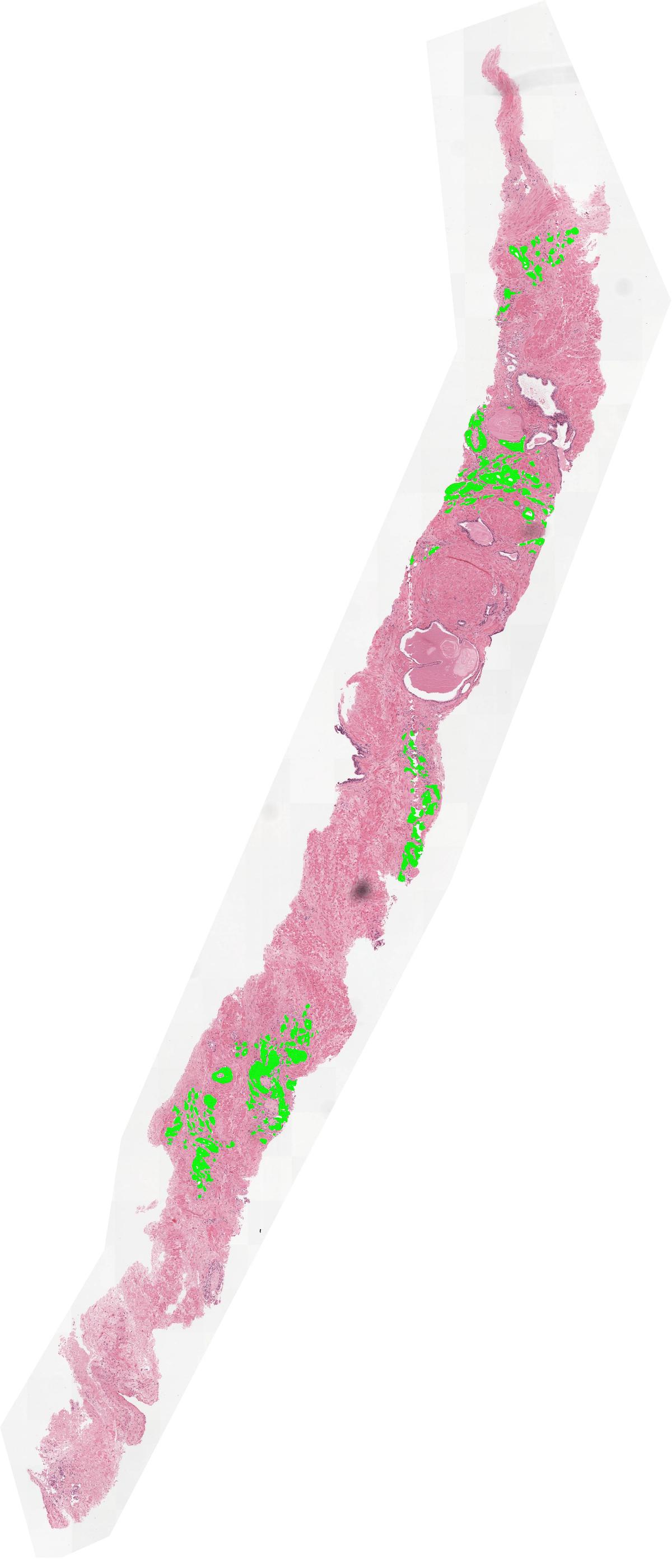} }
    \subfloat[]{\includegraphics[angle=90, width=0.5\textwidth]{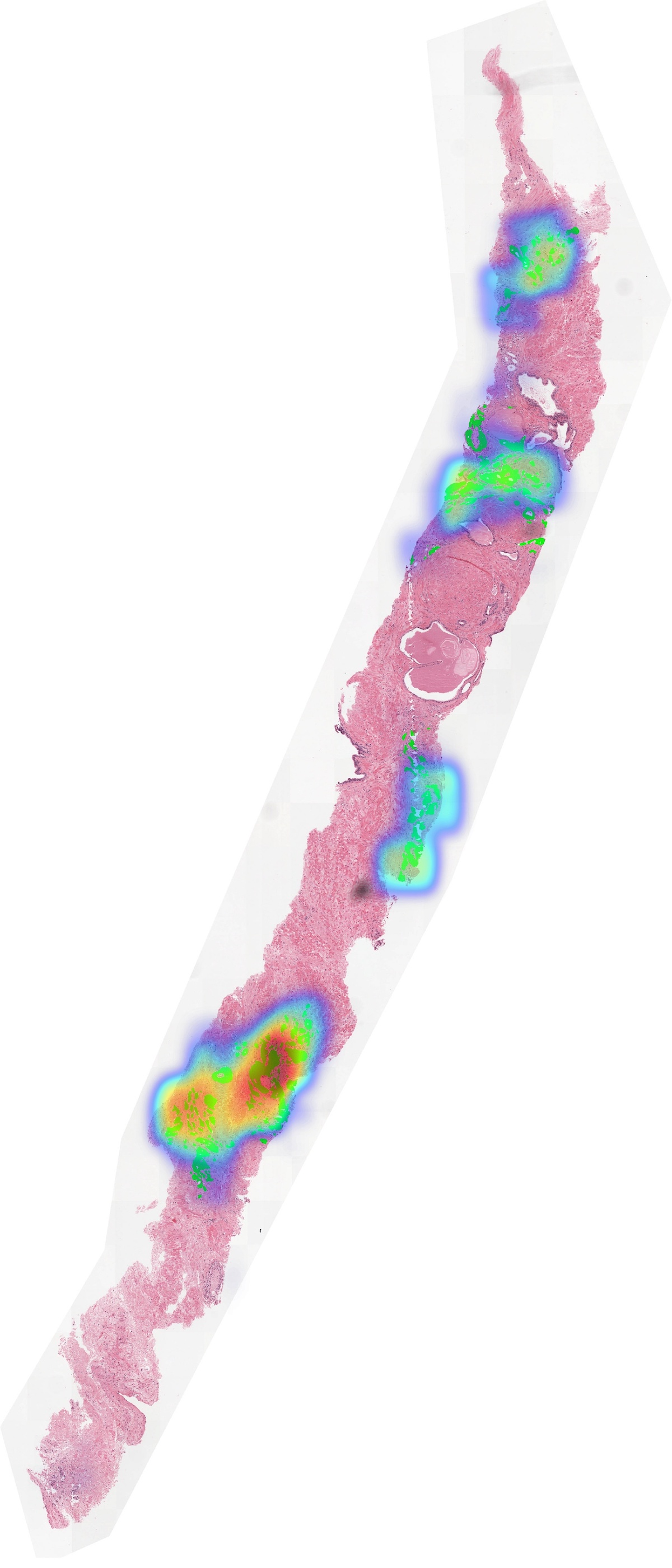}}%
    \caption{An example ISUP grade 5 prostate cancer WSI. (a) Green mask overlay shows ground truth location of cancer regions (provided  in the PANDA dataset~\cite{Panda2020}). (b) an additional heat map overlay visualizes our pseudo-labels of ISUP 5 (weighted by attention), achieved from training on weak (bag-level) labels only. Notice the close agreement between the dense pseudo-labels and the ground truth. In practice, pseudo-labels are computed per patch; here we used a sliding-window approach for dense visualization.}
    \label{fig:overlay}
\end{figure}

One of the challenges of MIL training is the lack of instance labels to guide the optimization process. A somewhat similar issue is encountered in semi-supervised learning~\cite{Xie2020},
where pseudo-labels are used  either offline or on the fly based on some intermediate estimates or another network's predictions. Here, we propose to generate pseudo-labels for each image patch and use the additional patch-wise loss to assist the optimization process.

\begin{equation}
\label{eq:add_loss}
L = L_{bag} +  \lambda \sum_k L_{patch}
\end{equation}
where the total loss $L$ includes a bag-level loss $L_{bag}$ (based on the ground truth labels) and a patch level loss $L_{patch}$ (based on the pseudo-labels). We use cross-entropy loss function for both bag-level and patch-level losses. 

We opt for a simple approach to generate pseudo-labels based on ensembling of several identical models trained from random initialization. The final ensembled labels are hard label (rounded to the  nearest classes). Consider a trained network, its bag-level prediction output is based on the final output vector $z$ (see Eq.~\ref{eq:weighted_sum}), followed by a linear projection onto the number of output classes: 
\begin{equation}
\label{eq:final_sum}
\c  = sigm(\W\z) = sigm(\W\HH\aa)
\end{equation}
here we assumed a final sigmoid function (but the same holds with softmax). We approximate the individual  instance level prediction as 
\begin{equation}
\label{eq:ind_pred}
\c_k = sigm(\W\h_k)
\end{equation}

\begin{algorithm}[t]
\SetAlgoLined
\DontPrintSemicolon
\SetAlgorithmName{Pseudocode}{} 
\KwResult{Train N MIL models (N=5)} \;
\For{all patches in the bag}{
  Run inference on patches for each image\;
  Ensemble predictions of attention weights $\aa$ and instance  classes $\c_k$\;
  \eIf{bag label is not zero}{
   for patches with top 10\% of highest attention $\aa$ weights, assign the ensembled labels as pseudo-label\;
   for patches with top 10\% of lowest attention $\aa$ weights, assign the zero labels as the pseudo-label\;
   otherwise flag the patch as unknown pseudo-label\;
   }{
   assign zero pseudo-labels for all patches, since here we know that all patches must have zero labels\;
  }
 }
 \caption{Pseudo-labels assignment}
 \label{alg:psuedolabels}
\end{algorithm}

Pseudocode~\ref{alg:psuedolabels} shows the  algorithm to compute the pseudo-labels. For some patches, whose ensembled attention weights are neither small nor large (defined by 10\% threshold), we do not assign any pseudo-labels, and mark then and unknown to exclude from the $L_{patch}$ loss. 
Given the pseudo-labels we re-optimize the model using the additional patch-wise loss.  The 10\% heuristic was chosen to retain only most confident patches, that contribute the most to the final bag-level classification.  A relevant approach was recently proposed by Lerousseau et al.~\cite{Lerousseau20}.  However the goal of their work is a dense segmentation map, and not the improvements to the global classification accuracy, and the pseudo-labels are calculated differently, through thresholding of current prediction probability estimates on the fly.

 \section{Experiments}
 \label{sec:experiments}
We implemented our method in PyTorch~\footnote{https://pytorch.org/} and trained it on  4 NVIDIA Tesla V100 16GB GPUs, batch size of 16. For the classification backbone, we use ResNet50 pretrained on ImageNet~\cite{He16}.  For the transformer layers, we keep a similar configuration as in~\cite{Vaswani17}, with 4 stacked transformer encoder  blocks. The lower pyramid level transformer has dimensionality of 256 for both input and hidden. The final transformer encoder has input dimension of 2308 (a concatenation of ResNet50 output features and the previous transformer outputs).  We use Adam optimizer with initial learning rate of $\alpha_{0}=3e-4$ for CNN parameters, and $3e-5$ for transformer parameters, then gradually decrease it using cosine learning rate scheduler for 50 epochs. We use 5-fold cross validations to tune the parameters. For transformer layers only, we use weight decay of $0.1$ and no dropout.

\subsubsection{PANDA dataset}
Prostate cANcer graDe Assessment (PANDA) challenge dataset consists of ~11K whole-slide images from two centers~\cite{Panda2020}. Currently, this is the largest public WSI dataset available. The grading process consisted of finding and classifying cancer tissue into Gleason patterns based on the architectural growth patterns of the tumor~\cite{Bulten21}. Consequently, it is converted into an ISUP grade on a 1-5 scale, based on the presence of two distinct Gleason patterns. The dataset was provided as part of the Panda kaggle challenge, which attracted more than 1000 teams, with the goal to predict the most accurate ISUP grades. 
Each individual image on average is about 25,000px$\times$25,000px RGB. 
The challenge also includes a hidden dataset, whose images were graded by multiple pathologists. The private dataset labels are not publicly available, but can be used to asses your model blindly via Kaggle website (invisible to the public as the challenge is closed now). In our experiments, we use a medium resolution input images (4x smaller than the highest resolution).

\subsubsection{Patch selection}
To extract patches from WSI, we tile the the image into a grid of 224px$\times$224px patches. At each iteration, the grid has a random offset from the top left corner, to ensure randomness of the patches. We then retain only the foreground patches. 
From the remaining patches, we maintain only a random subset (K=56), which is a trade-off between covering the tissue content and GPU memory limits (see Figure~\ref{fig:patches}). We use batch size 16, which makes the data input size $16\times K\times 3\times 224\times 224$ at each iteration. During testing, inference is done using all foreground patches.

\subsection{Results}
\subsubsection{Transformer MIL} 
We evaluate and compare our method to the Attention MIL and its Gated Attention MIL~\cite{Ilse18}, as well as to a classical MIL with Max operator~\cite{Campanella19}. 
For evaluation metrics we use Accuracy, Area Under Curve (AUC) and Quadratic Weighted Kappa (QWK) of ISUP grade prediction (see Table~\ref{tab:panda1}). QWK metric measures the similarity between the predictions and targets, with a maximum value of 1. QWK was chosen as the main metric during the PANDA challenge~\cite{Panda2020}, since it is more appropriate for the tasks with predicted classes being severity grades/levels. All metrics are computed using our 5-fold (80\%/20\%  training/validation) splits, except for the \emph{Leaderboard} column results, which come from the evaluation on kaggle challenge hidden private test-set. Even though the challenge is closed now, it allows for blind submission of the code snippet, which runs on the  PANDA hidden set and outputs the final QWK number. These results are not added to the kaggle leaderboard, and are allowed only for post-challenge evaluations. Table~\ref{tab:panda1} shows that  the proposed two transforms based approaches outperform other methods both in our validation sets, and on the challenge hidden set.  We have also inspected the self-attention matrices and found that for many cases, they have have distinct off-diagonal high value elements.
In particular, instances with WSI tumor cells of different Gleason scores have higher off-diagonal values, indicating that such a combination is valuable for the final classification, which was captured by the transformer self-attention.

\subsubsection{Patch-wise pseudo-labels} 

We train 5 models and ensemble their patch-level predictions. We use $\lambda=100$.  We show the performance of adding the pseudo-labels supervision in Table~\ref{tab:pseudo1}. In all cases the performance has improved compared to the baselines shown in Table~\ref{tab:panda1} by $\sim1\%$. Table~\ref{tab:pseudo1} also shows the QWK results of the winners (top 3 places) of the PANDA kaggle challenge. Notice that our single model results are on par with the winners of the challenge (who all use ensembling of several models). We also experimented with ensembling, and the ensemble of our 10 models, achieves the leaderboard QWK of 0.94136, which would have been the first place in the leaderboard.

We have also tried but found no benefit of repeating pseudo-labeling several rounds, because the pseudo-label values almost do not change after the 1st round.  

\begin{table}[ht]
	\centering
	\caption{Evaluations results on PANDA dataset. The Leaderboard column shows the QWK results of the private leaderboard of Kaggle's challenge, which allows direct comparison to more then 1000 participants.}
	\label{tab:panda1}
	\begin{tabular}{|l|c|c|c|c|}

		\hline
		 &  Accuracy & AUC &  QWK  & Leaderboard \\ \hline
		Attention MIL~\cite{Ilse18}           & $0.793\pm0.035$ & $0.983\pm0.021$   & $0.948\pm0.036$ & $0.915\pm0.086$ \\
		Gated attention MIL~\cite{Ilse18}      & $0.795\pm0.037$ & $0.981\pm0.011$  & $0.936\pm0.042$ & $0.914\pm0.069$  \\
		Max MIL~\cite{Campanella19}                  & $0.770\pm0.055$ & $0.973\pm0.048$    & $0.910\pm0.053$ & $0.868\pm0.091$  \\
		\hline
		Transformer MIL        & $0.801\pm0.014$ & $0.988\pm0.015$    & $0.960\pm0.034$ & $0.930\pm0.012$  \\
		Pyramid Transformer MIL & $\mathbf{0.805}\pm0.011$ & $\mathbf{0.989}\pm0.018$    & $\mathbf{0.961}\pm0.032$ & $\mathbf{0.932}\pm0.015$  \\
		\hline

	\end{tabular}

\end{table}
\begin{table}[h]
	\centering
	\caption{Evaluation results of adding pseudo-labels to our baseline transformer MIL approaches. 
	We also include the results of the top three places of this challenge\protect\footnotemark (who all use ensembling of several models). Our results indicate that pseudo-labeling further improves the performance, with our single model providing results on par with the top winning teams.}
	
	\label{tab:pseudo1}
	\begin{tabular}{|l|c|c|}

		\hline
		 &   QWK (val)  & QWK (Leaderboard) \\ \hline
		Attention MIL~\cite{Ilse18} + Pseudo-labels       & $0.9502\pm0.0319$ & $0.9304\pm0.0542$ \\
		\hline
		Transformer MIL + Pseudo-labels     & $0.9614\pm0.0367$ & $0.9347\pm0.0353$ \\
		Pyramid Transformer MIL + Pseudo-labels   & $\mathbf{0.9652}\pm0.0168$ & $\mathbf{0.9365}\pm0.0513$  \\
		\hline
		First place - Panda kaggle challenge~\cite{Panda2020}    & - & $0.94085$  \\
		Second place - Panda kaggle challenge~\cite{Panda2020}    & - & $0.93768$  \\
		Third place  - Panda kaggle  challenge~\cite{Panda2020}    & - & $0.93480$  \\
		Pyramid Transformer MIL (ours, ensemble of 10)    & - & $\mathbf{0.94136}$  \\
		\hline
	\end{tabular}
\end{table}

\footnotetext{https://www.kaggle.com/c/prostate-cancer-grade-assessment/leaderboard}

\section{Discussion and Conclusion}
 \label{sec:conclusion}
 
We proposed a new deep learning based MIL approach for WSI classification with the  following two main contributions: the addition of the transformer  module to account for dependencies among instances and the instance-level supervision loss using pseudo-labels. We evaluated the method on PANDA challenge prostate WSI dataset, which includes over 11000 images. To put in perspective, most recently published SOTA methods evaluated their performance on datasets with the order of only several hundred images~\cite{Zhao20,Hashimoto20,Lu21,Maksoud20}.
Furthermore, we compared our results directly to the leaderboard of the PANDA kaggle challenge with over 1000 participating teams, and demonstrated that our single model performance is on par with the top three winning teams, as evaluated blindly on the same hidden private test-set. Finally, recently proposed visual transformers~\cite{Dosovitskiy21} have shown a capability to replace the classification CNN completely, allowing for the possibility to create deep learning based MIL model solely based on the transformer blocks; we leave these investigations for future research.

\bibliographystyle{splncs04}
\bibliography{paper}

\end{document}